\documentclass[aps,showpacs,preprint]{revtex4-1}
\usepackage{graphicx}
\usepackage{color}

\usepackage[table]{xcolor}
\usepackage{ctable}
\usepackage[T1]{fontenc}
\usepackage[latin9]{inputenc}
\usepackage{amsmath}
\usepackage{amssymb}

\begin{document}

\title{Experimental simulation of quantum tunneling in small systems\footnote{Feng, G., Lu, Y., Hao, L., Zhang, F. \& Long, G. Experimental
simulation of quantum tunneling in small systems. Sci. Rep. 3, 2232; DOI:10.1038/srep02232 (2013).}}
\author{Guan-Ru Feng$^{1,2,\dag}$, Yao Lu$^{1,2,\dag}$, Liang Hao$^{1,2}$,
Fei-Hao Zhang$^{1,2}$ \& Gui-Lu Long$^{1,2}$
\footnote{Correspondence  to gllong@tsinghua.edu.cn. $\dag $  These authors contributed equally to this work. }
}

\newcommand{\mm}[1]{\Large $\mathbf{#1}$}

\address{ State Key Laboratory of Low-dimensional Quantum Physics and
Department of Physics, Tsinghua University, Beijing 100084, P. R.
China.
\\
Tsinghua National Laboratory for Information Science and
Technology,  Beijing 100084, P. R. China.}

\begin{abstract}
It is well known that quantum computers are superior to classical
computers in efficiently simulating quantum systems. Here we report
the first experimental simulation of quantum tunneling through
potential barriers, a widespread phenomenon of a unique quantum
nature, via NMR techniques. Our experiment is based on a digital
particle simulation algorithm and requires very few spin-1/2 nuclei without the need of
ancillary qubits. The occurrence of quantum tunneling through a
barrier, together with the oscillation of the state in potential wells, are clearly observed through the experimental results. This experiment has clearly demonstrated the possibility to observe and study profound physical phenomena within even the reach of small quantum computers.
\end{abstract}

\maketitle

Quantum simulation is one of the most important aims of quantum computation ever since Feynman studied the likelihood of simulating one quantum system by another \cite{Feynman}. Recent years
have witnessed fruitful results in the development of quantum
computation, and it has been demonstrated that quantum computers
can solve certain types of problems with a level of efficiency beyond the
capability of classical computers \cite{shor,Grover,Simon,walk,DJ}, among which the simulation of the dynamics of quantum systems is especially attractive because of the exponential improvement in computational resources and speeds.  Quantum simulation has become a subject of intense investigation and has been realized in various situations, such as system evolution with a
many-body interaction Hamiltonian \cite{Tseng,Negrevergne,Peng1,fenggr}, the
dynamics of entanglement \cite{Zhang,Peng}, quantum phase
transitions \cite{Edwards,Zhangphase}, and calculations of molecular
properties \cite{Lanyon,Du,Yung,Aspuru,Kassal}.

Quantum tunneling plays an essential role in many quantum phenomena, such as
the tunneling of superconducting Cooper pairs \cite{Josephson} and alpha decay \cite{Gamow}. Moreover, tunneling has been widely applied in modern devices and modern experimental techniques, such as the tunnel diode \cite{Esaki}, the scanning tunneling
microscope \cite{Binnig} and so on. As a unique fundamental concept in quantum mechanics, the simulation of quantum tunneling is of great significance. Many important science problems, such as lattice quantum chromodynamics \cite{lqcd}, can be dealt with similarly. However due to the large number of quantum gates and qubits required, the simulation of quantum tunneling in a quantum computer has remained untested experimentally. Recently Sornborger \cite{Andrew} proposed a
digital simulation algorithm for demonstrating the tunneling of a particle in a double-well potential with no ancillary qubits, and at least halved the number of quantum gates. This makes it possible to simulate this important quantum effect in today's quantum information processors with only a few qubits.
 In this paper, we report the first experimental digital quantum simulation of this significant quantum phenomenon via a liquid nuclear magnetic resonance (NMR) quantum information processor. In the experiment, the continuous process of one-dimensional
tunneling of a particle through a potential barrier is clearly
demonstrated, and the oscillation of the particle in potential wells is clearly observed. Our experiment has shown that with very few qubits, interesting quantum effects such as tunneling dynamics can be simulated with techniques which are within reach of current quantum architectures.\\

\section*{Results}
{\bf Theoretical protocol.} Consider  a single particle moving
in a square well potential in one-dimensional space. The  Schr\"{o}dinger equation reads,
\begin{equation}
i\frac{\partial}{\partial t}\left|\psi\left(x,t\right)\right\rangle =\left[\frac{\hat{P}^{2}}{2m}+V(\hat{X})\right]\left|\psi\left(x,t\right)\right\rangle ,\label{schrodinger}\end{equation}
 where $\hat{P}$ and $\hat{X}$ are momentum and position operators,
respectively. Throughout the text we set $\hbar$ to 1. The evolution of the wave function with time can
be straightforwardly given as
 \begin{equation}
\left|\psi\left(x,t+\Delta t\right)\right\rangle =e^{-i\left[\frac{\hat{P}^{2}}{2m}+V(\hat{X})\right]\Delta t}\left|\psi\left(x,t\right)\right\rangle .\label{def-1}\end{equation}

In the digital quantum simulation \cite{Zalka,Benenti,Andrew}, the continuous  coordinate $x$ is discretized. Suppose $\psi\left(x,t\right)$ is continuous on the region $0<x<L$,
and with a periodic boundary condition $\psi\left(x+L,t\right)=\psi\left(x,t\right)$.
$x$ is discretized on a lattice with spacing $\Delta l$ and the
wave function is stored in an $n$-qubit quantum register \begin{equation}
\left|\psi\left(x,t\right)\right\rangle \rightarrow\sum_{k=0}^{2^{n}-1}\psi\left(x_{k},t\right)\left|k\right\rangle ,\label{register}\end{equation}
 where $x_{k}=(k+\frac{1}{2})\Delta l$, $\Delta l=\frac{L}{2^{n}}$,
and $\left|k\right\rangle $ is the lattice basis state corresponding
to the binary representation of number $k$. Equation (\ref{register})
gives a good approximation to the wave function in the limit $n\rightarrow\infty$.

In the lattice space of the digital quantum simulation, an important task is to construct the kinetic and potential
operators. Because the potential operator
$V(\hat{X})$ is a function of the coordinate operator $\hat{X}$, it is a diagonal matrix in the coordinate representation, which can be decomposed as \cite{Bullock} \begin{equation}
V=\sum_{i_{1},i_{2},...,i_{n}=3}^{4}c_{i_{1}i_{2}...i_{n}}\otimes_{k=1}^{n}\sigma_{i_{k}},\label{potential}\end{equation}
where $\sigma_{3}$ is the Pauli
matrix $\sigma_{z}$, and $\sigma_{4}=I$ is the identity matrix in two dimensions.

The kinetic energy operator, which is diagonal in the momentum representation,
can be constructed in the coordinate representation with the help of a quantum Fourier transformation (QFT).
Similar to equation (\ref{register}), we have \begin{equation}
\left|\phi\left(p,t\right)\right\rangle \rightarrow\sum_{j=0}^{2^{n}-1}\phi\left(p_{j},t\right)\left|j\right\rangle ,\label{pregister}\end{equation}
 where $\phi\left(p,t\right)$ is the wave function in the momentum representation, and

\begin{eqnarray}
p_{j}=\left\{
\begin{array}{cc} 2\pi j/2^{n},           & 0\leqslant j\leqslant2^{n-1},\\
                       {2\pi}(2^{n-1}-j)/2^{n}, & 2^{n-1}<j<2^{n},\\
                       \end{array}\right.
\end{eqnarray}
is the eigenvalue of the momentum operator $\hat{P_{p}}$ in  the momentum representation,
\begin{equation}
\hat{P_{p}}=\sum_{j=0}^{2^{n-1}}\frac{2\pi}{2^{n}}j\left|j\right\rangle \left\langle j\right|+\sum_{j=2^{n-1}+1}^{2^{n}-1}\frac{2\pi}{2^{n}}(2^{n-1}-j)\left|j\right\rangle \left\langle j\right|.\end{equation}
 Therefore
the kinetic energy operator in the coordinate representation can be
written via the QFT as \begin{equation}
\frac{\hat{P^{2}}}{2m}=\textbf{F}^{-1}\frac{\hat{P_{p}^{2}}}{2m}\textbf{F},\label{fourier}\end{equation}
 where  \begin{eqnarray}
\textbf{F}=\frac{1}{2}\sum_{j,k=0}^{2^n-1}e^{\frac{1}{2}i\pi jk}\left|j\right\rangle \left\langle k\right|,\label{fourier3}\end{eqnarray}
is the QFT operator.

Equations (\ref{register}), (\ref{potential}) and (\ref{fourier}) give the discretized forms of the wave function, the potential operator
and the kinetic energy operator, respectively.
With these expressions in hand,  we can efficiently implement the time evolution of the system within a small interval $\Delta t$, by using a modified
Trotter formula, which is correct up to $\Delta t$ \cite{Zalka,Andrew,Nielsen}.   Equation (\ref{def-1}) is then approximated as
\begin{equation}
\left|\psi\left(x,t+\Delta t\right)\right\rangle =e^{-i\frac{\hat{P}^{2}}{2m}\Delta t}e^{-iV(\hat{X})\Delta t}\left|\psi\left(x,t\right)\right\rangle .\label{Trotter
2}\end{equation}
From equation (\ref{fourier}) we have
\begin{equation}e^{-i\frac{\hat{P}^{2}}{2m}\Delta t} =\textbf{F}^{-1}e^{-i\frac{\hat{P}_{p}^{2}}{2m}\Delta t} \textbf{F}=F^{-1}DF,\label{kinetic6}\end{equation}
where $F$ is the usual bit-swapped Fourier transformation operator, which can be
readily realized in quantum circuits via a series of Hadamard gates
and controlled-phase gates \cite{Coppersmith}. Consequently $D$ in equation (\ref{kinetic6}) is a bit-swapped version of $\exp\left[-i\hat{P}_{p}^{2}\Delta t/{(2m)}\right]$.
Thus equation (\ref{Trotter 2}) can be rewritten as \begin{equation}
\sum_{k=0}^{2^n-1}\psi\left(x_{k},t+\Delta t\right)\left|k\right\rangle =F^{-1}DFQ\sum_{k=0}^{2^n-1}\psi\left(x_{k},t\right)\left|k\right\rangle ,\label{Trotter 3}\end{equation}
where $Q=\exp[-iV(\hat{X})\Delta t]$.

From equation (\ref{Trotter 3}) the one time step evolution quantum
circuit is straightforwardly obtained. In Figs. 1 and 2 we draw the two-qubit and the three-qubit implementations, respectively. The explicit construction of $F$, $D$, and $Q$ is detailed in Methods.

{\bf Experimental procedures and results.} As a small-scale demonstration, a two-qubit simulation and a three-qubit simulation were investigated. In the experiment, we studied the time evolution of a particle moving in double-well potentials using the digital algorithm via two-qubit and three-qubit NMR quantum information processors. The molecular structures and parameters of the quantum information processors are given in Fig. 3.

In the two-qubit case, the four basis states $|00\rangle$, $|01\rangle$, $|10\rangle$ and $|11\rangle$ register the four lattice sites 1, 2, 3 and 4 as discretized position variables of the particle. We consider the potential $V_{0}I\otimes\sigma_{z}$, which can be implemented with only a single-qubit gate \cite{Andrew}. This potential represents a double-well potential of amplitude $2V_{0}$, with two peaks $V_{0}$ at $|00\rangle$ and $|10\rangle$, and two troughs $-V_{0}$ at $|01\rangle$ and $|11\rangle$. In our experiments we set the time interval   $\Delta t=0.1$ and the amplitude of the potential $2V_{0}=20$. We
simulated the situation in which the particle is initially trapped
inside one of the two wells by preparing the pseudo-pure state
$|01\rangle$ from thermal equilibrium \cite{pps} as the initial
state.

The tunneling process of the particle was simulated using nine
experiments, where in each experiment the operations in
Fig. 1 were performed.  Quantum
state tomography (QST) was performed on the density matrices of the
final states after 1 to 9 steps (\cite{tomography1,tomography2}).
The whole evolution process   of a single
particle in the double-well potential can
be described by depicting the diagonal elements of the density matrix, which correspond to the probability distribution of the particle, at each step. They are illustrated in Fig. 4 (b).  It is clearly
observed that the particle tunnels through the potential barrier
between the two wells while its probability to be found in the
barrier remains scarce, which accords well with theoretical
calculations in Fig. 4 (a).

For the sake of comparison we also experimentally simulated the
evolution of a free particle with zero potential using
the same experimental schemes and parameters, except the removal
of the potential operator $Q$ in the circuit. The experimental results, together with theoretical
calculations, are plotted in Fig. 4 (c) and (d). It is not surprising to find that
the probability is distributed more evenly on all the four sites, showing the particle is free.

Similarly, in the three-qubit case, the eight basis states $|000\rangle$, $|001\rangle$,
$|010\rangle$, $|011\rangle$, $|100\rangle$, $|101\rangle$, $|110\rangle$ and $|111\rangle$ register the eight lattice sites 1, 2, 3, 4, 5, 6, 7 and 8.
The potential considered here is $V_{0}I\otimes\sigma_{z}\otimes I$, which corresponds to a double-well potential of amplitude $2V_{0}$. It has a higher resolution than the  double-well potential in the two-qubit simulation, with each well or peak occupying two sites: one well occupying $|010\rangle$ and $|011\rangle$, and the other well occupying $|110\rangle$ and $|111\rangle$. In our experiments we set the time interval  to  $\Delta t=0.4$ and the amplitude of the potential $2V_{0}=200$. The initial position of the particle is set in one of the two wells, by preparing the pseudo-pure state
$|110\rangle$ as the initial state.

Five experiments have been carried out, where the evolution time is $n\Delta t$, with $n$ being in 1,2,3,4 and 5, respectively. The reconstructed diagonal elements of the density matrices, which correspond to the probability distribution of the particle at each site, are illustrated in Fig. 5 (b). Not only the  particle's tunneling from one well to the other,  but also its oscillations within each well, disappearing and then appearing again,  are clearly observed from Fig. 5 (b), which are in good agreement with theoretical calculations in Fig. 5 (a).\\

\section*{Discussion}

Our experiments have simulated the fundamental quantum phenomenon of tunneling. Our  two-qubit experiments reflected a remarkable difference between the two situations with and without the double-well potential. In our three-qubit experiments, a higher resolution can provide a more sophisticated structure of the potential wells, making it possible to observe the in-well-oscillation of the particle.

In the two-qubit simulation, the experimental density matrices for the initial state $|01\rangle$ and the final state after 9 time steps were fully reconstructed (see Fig. 6) with experimental fidelities 99.89\% and  95.48\%, respectively. In the three-qubit simulation, the experimental density matrices for the initial state $|110\rangle$ and the final state after 5 time steps were also fully reconstructed (see Fig. 7) and the experimental fidelities for them are 98.63\% and  93.81\%, respectively. The high fidelities demonstrate our good control in the experiments.

It should be emphasized here that although the real evolution of the particle takes place in a continuous space with infinite dimensions, our quantum computer, which works with only a few qubits in limited dimensions, is already capable of undertaking some basic
yet fundamental simulation tasks, such as quantum tunneling. Likewise, an $N$-site-lattice simulation can be efficiently  implemented in experiments with only ${\rm log}_{2}N$ qubits. The
result has revealed the amazing power hidden behind the qubits and
the promising future of quantum simulations.

In summary, we accomplished a small-scale demonstration of the
quantum tunneling process on two- and three-qubit NMR systems based on the
digital quantum simulation method. This is the first experimental digital quantum simulation of quantum tunneling  via NMR quantum information processors. The experimental results and
the theoretical predictions are in good agreement.\\

{\noindent \bf Methods}

 Experiments were carried out at room temperature using a Bruker Avance III 400 MHz spectrometer. We used chloroform dissolved in $d6$ acetone as the two-qubit NMR quantum processor, and diethyl-fluoromalonate dissolved in $d6$ acetone as the three-qubit NMR quantum information processor. The natural Hamiltonians of the two-qubit system (denoted as $H_{\rm NMR}^{2}$) and the three-qubit system (denoted as $H_{\rm NMR}^{3}$) are as follows,
\begin{align}
H_{\rm NMR}^{2}=&-\Sigma_{i=1}^{2}\pi \nu_{i}\sigma_{z}^{i}+\frac{\pi}{2}J_{1,2}\sigma_{z}^{1}\sigma_{z}^{2},\\
H_{\rm NMR}^{3}=&-\Sigma_{i=1}^{3}\pi \nu_{i}\sigma_{z}^{i}+\Sigma_{i<j,i=1}^{3}\frac{\pi}{2}J_{i,j}\sigma_{z}^{i}\sigma_{z}^{j},
\end{align}
where $\nu_{i}$ is the chemical shift of the $i$th nucleus, and $J_{i,j}$ is the $J$-coupling constant between the $i$th and the $j$th nuclei. The molecular structures of chloroform and diethyl-fluoromalonate, and their parameters are described in Fig. 3. The two quantum information processors are initially prepared in the pseudo-pure states $|01\rangle $ and $|110\rangle$ using the spatial average technique \cite{pps}, with qubit orders ${\rm^{1}H, ^{13}C}$ and ${\rm^{1}H, ^{13}C, ^{19}F}$, respectively.

{\bf Gate construction and implementation in the two-qubit simulation.} Here we give the explicit construction of $F$, $Q$,  and $D$, which are illustrated in Fig. 1. $F$ is the bit-swapped Fourier transformation operator, which can be implemented in quantum circuits via a series of Hadamard gates
and controlled-phase gates \cite{Coppersmith}, $
F=H_{2}R_{\frac{\pi}{2}}H_{1}$, where $H_{1}$
and $H_{2}$ are Hadamard gates on the first and second qubits, respectively, and $R_{\frac{\pi}{2}}={\rm diag}\;[1,1,1,i]$ is the
two-qubit controlled-phase gate. $Q=e^{-iV(\hat{X})\Delta t}=I\otimes e^{-iV_{0}\sigma_{z}\Delta t}$ is a single-qubit gate on the second qubit. $D$ realizes a bit-swapped version of $\exp[-i\hat{P}_{p}^{2}\Delta t/(2m)]$ up to an overall phase factor (we have taken $m$ = 1/2), as $D=\Phi_{\pi}Z_{1}Z_{2}$, where $\Phi_{\pi}$, $Z_{1}$, and $Z_{2}$ can be expressed as follows,
\begin{align}
\Phi_{\pi}=&e^{-i\frac{\pi}{2}^{2}(R_{\frac{\pi}{2}})^2\Delta t},\nonumber\\
Z_{1}=&e^{i\frac{\pi}{8}^{2}\sigma_{z}\otimes I\Delta t},\nonumber\\
Z_{2}=&e^{i\frac{\pi}{2}^{2}I\otimes\sigma_{z}\Delta t}.\nonumber\end{align}

The single-qubit gates in Fig. 1 are realized using radio-frequency pulses, and the two-qubit gates in Fig. 1 are realized by combining refocusing pulses and $J$-coupling evolution. The pulse sequences for implementing $F$, $D$ and $Q$ are exhibited in Fig. 8.

{\bf Gate construction and implementation in the three-qubit simulation.}
The operations $H_{i}$ ($i=1,2,3$) in Fig. 2 are Hadamard gates on the $i$th qubit, and $R_{\frac{\pi}{4}}={\rm diag}\;[1,1,1,\exp(i\frac{\pi}{4})]$. The other operations in Fig. 2 are as follows:
\begin{align}
Q&=e^{-iV(\hat{X})\Delta t}=I\otimes e^{-iV_{0}\sigma_{z}\Delta t}\otimes I,\nonumber\\
Z_{1}&=e^{i\frac{\pi}{32}^2\sigma_z\otimes I\otimes I\Delta t},\nonumber\\
Z_{2}&=e^{i\frac{\pi}{8}^2 I\otimes\sigma_z\otimes I\Delta t},\nonumber\\
Z_{3}&=e^{i\frac{\pi}{2}^2 I\otimes I\otimes\sigma_z\Delta t},\nonumber\\
\Phi_{1}&=e^{-i\frac{\pi}{2}^2 {\rm diag}[1,1,1,-1]_{23}\Delta t},\nonumber\\
\Phi_{2}&=e^{-i\frac{\pi}{4}^2{\rm diag}[1,1,1,-1]_{13}\Delta t},\nonumber\\
\Phi_{3}&=e^{i\frac{\pi}{8}^2 {\rm diag}[1,1,1,-1]_{12}\Delta t}.\nonumber
\end{align}

We used the optimal control method for pulse designing in our experiments, namely the GRAPE algorithm \cite{GRAPE}, which is widely used for designing and optimizing pulse sequences to produce a desired unitary propagator.  The evolution operators $\exp[-i(\textbf{F}^{-1}\hat{P}_{p}^{2}/{(2m)}\textbf{F}+V(\hat{X}))n\Delta t]$ , with $n$=1,2,3,4 and 5,  which are illustrated in Fig. 2, are implemented using optimized pulses generated by the GRAPE algorithm in the experiments.


{\noindent Addendum} 

This work was supported by the National Natural Science Foundation
of China (Grant Nos. 11175094, 91221205), the National Basic Research Program
of China (2009CB929402, 2011CB921602), and the Specialized Research
Fund for the Doctoral Program of Education Ministry of China.


G.R.F., Y.L., F.H.Z. and G.L.L. designed the experimental scheme; G.R.F., Y.L. and G.L.L. performed the experiments; G.R.F., L.H. and G.L.L. analyzed data; G.R.F., Y.L. and G.L.L. wrote the paper.


 Competing financial interests: The author declares no competing financial interests.

\begin{figure}[tbp]
\begin{center}
\includegraphics[width=88mm]{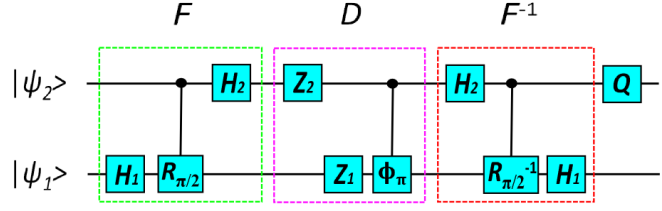}
\end{center}
\caption{
The two-qubit quantum circuit for one step of simulation. $|\psi_{1}\rangle$
and $|\psi_{2}\rangle$ are the input states of the first and the
second qubits, respectively.
}
\label{circuit}
\end{figure}

\begin{figure}[tbp]
\begin{center}
\includegraphics[width=88mm]{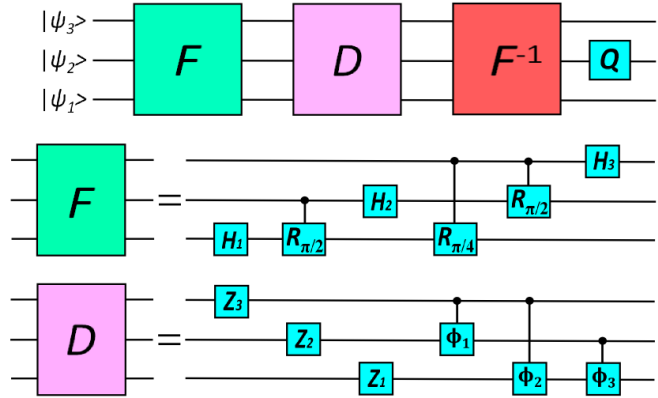}
\end{center}
\caption{
 he three-qubit quantum circuit for one step of simulation. $|\psi_{1}\rangle$, $|\psi_{2}\rangle$, and $|\psi_{3}\rangle$ are the input states of the first, the
second, and the third qubits, respectively.}
\label{circuit3}
\end{figure}

\begin{figure}[tbp]
\begin{center}
\includegraphics[width=88mm,height=72mm]{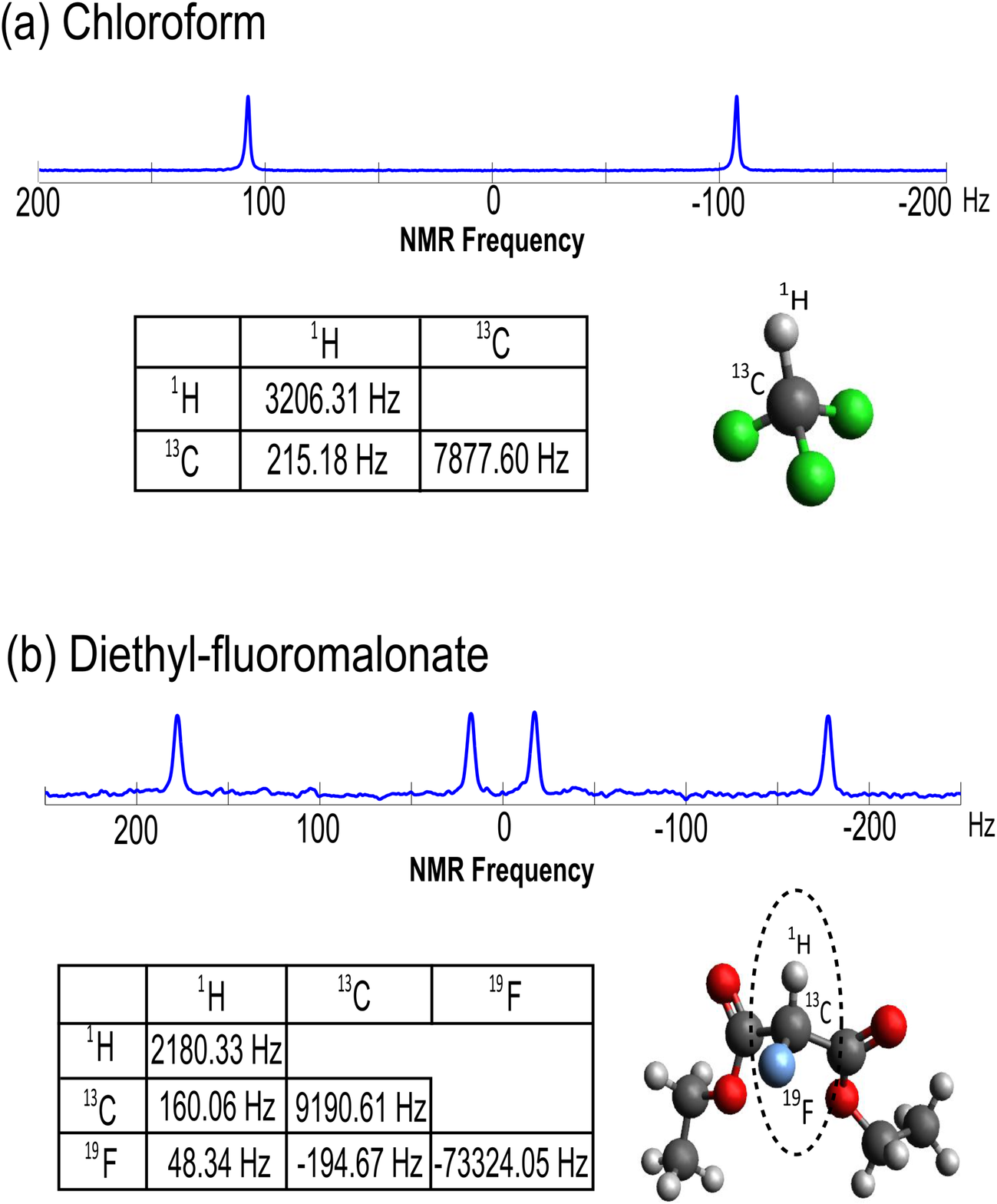}
\end{center}
\caption{
Molecular structure and properties of the NMR quantum information processors:  chloroform and diethyl-fluoromalonate. The chemical shifts and  $J$-coupling constants of the two molecules are given as diagonal elements and off-diagonal elements in the two tables, respectively. The chemical shifts are given with respect to reference frequencies of 400.13 MHz (hydrogens), 100.61 MHz (carbons) and 376.50 MHz (fluorines). $\rm{^{1}H}$ and $\rm{^{13}C}$ in the chloroform molecule act as qubits 1 and 2 in the two-qubit simulation experiment. $\rm{^{1}H}$, $\rm{^{13}C}$ and $\rm{^{19}F}$ in the diethyl-fluoromalonate molecule act as qubits 1, 2 and 3 in the three-qubit simulation experiment. The NMR spectra of the $\rm{^{13}C}$ obtained through a read-out $\frac{\pi}{2}$ pulse on the equilibrium states of chloroform and diethyl-fluoromalonate are also shown here.}
\label{moleculestructure}
\end{figure}

\begin{figure}[tbp]
\begin{center}
\includegraphics[width=180mm,height=100mm]{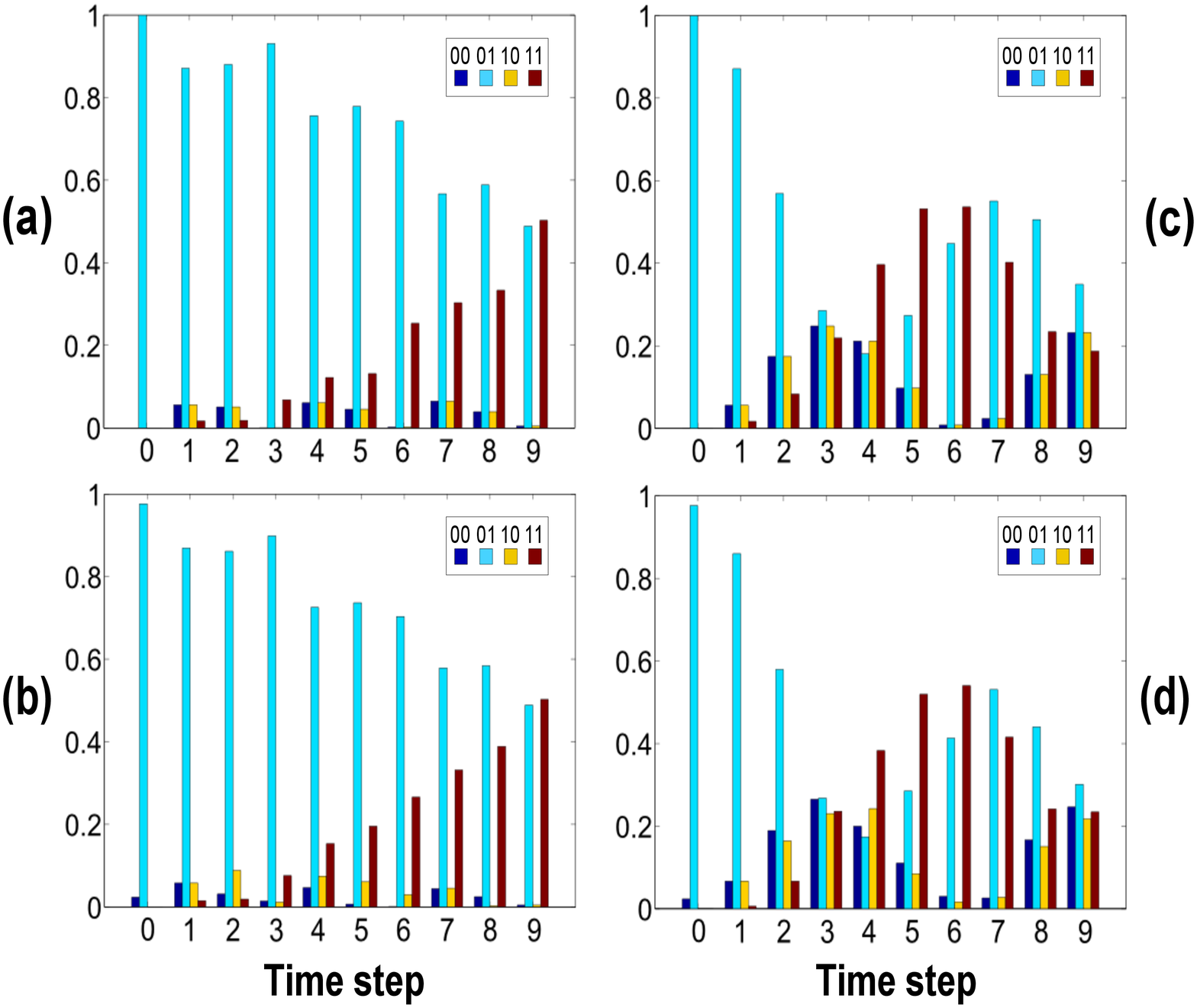}
\end{center}
\caption{
(a) and (b) illustrate a single particle's probability distribution as a
function of time for the nine steps for a particle in  a
double-well potential in the two-qubit simulation. The two potential wells are at $|01\rangle$
(site 2) and $|11\rangle$ (site 4).
 (c) and (d) illustrate the free particle's probability
 distribution as a function of time for the nine steps. (a) and (c) are the
theoretical predictions; (b) and (d) are the liquid NMR experimental results. The four bars at each step indicate the particle's probabilities on
the states $|00\rangle$ (blue), $|01\rangle$ (green), $|10\rangle$
(yellow), $|11\rangle$ (red), from left to right. The initial state
is prepared in $|01\rangle$. As is illustrated in (a) and (b), the amplitude of $|01\rangle$ decreases with time while that of $|11\rangle$ increases, clearly
manifesting the tunneling from site 2 to site 4. As for the free particle situation, the probability distribution becomes more even among all the
four states, as illustrated in (c) and (d). Our experimental results
agree well with theoretical calculations.}
\label{tunneling}
\end{figure}

\begin{figure}[tbp]
\begin{center}
\includegraphics[width=180mm,height=100mm]{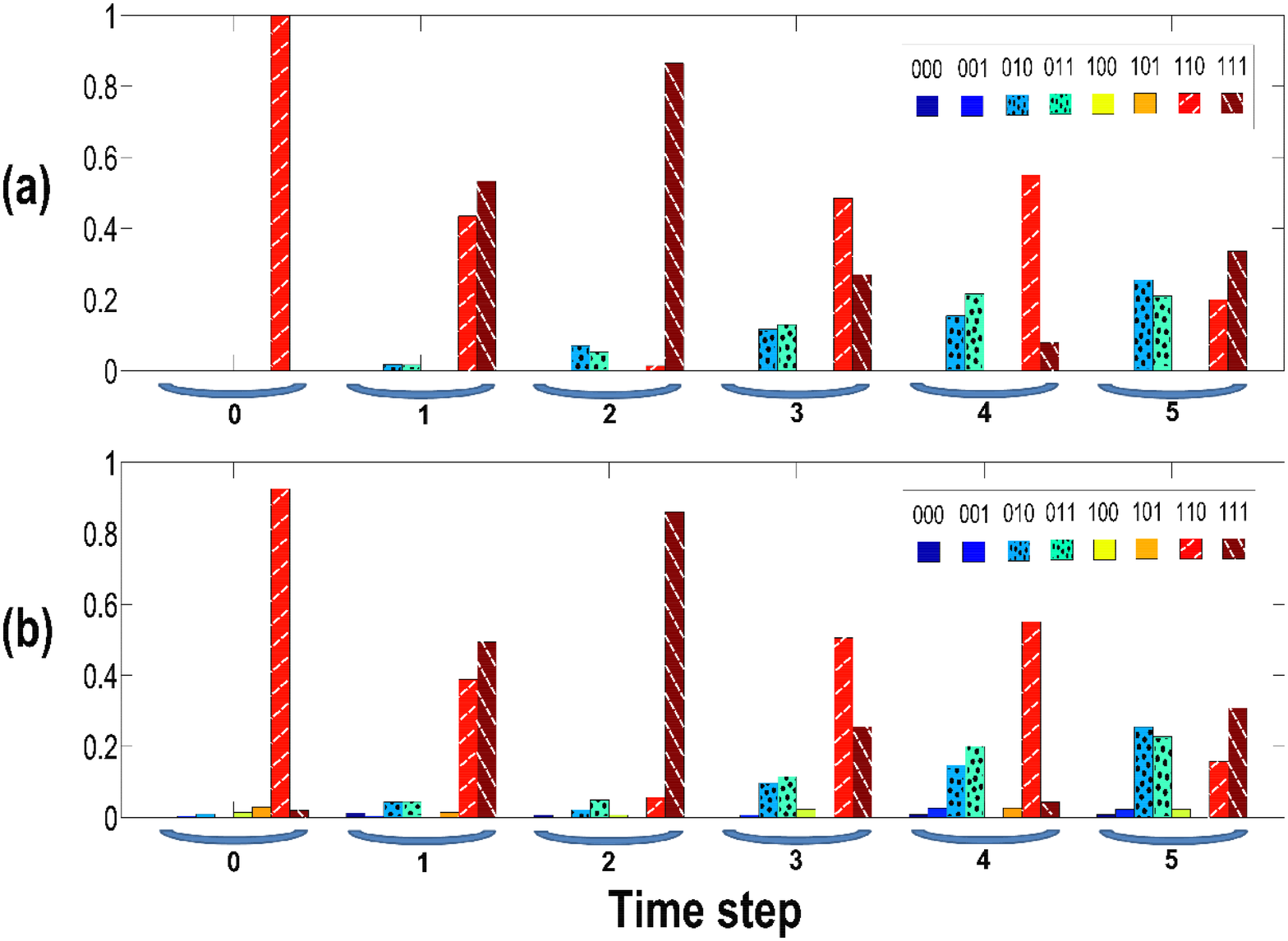}
\end{center}
\caption{
 A single particle's probability distribution as a
function of time for the five time steps for a particle in  a
double-well potential in the three-qubit simulation. One well is at $|010\rangle$
(site 3) and $|011\rangle$
(site 4). The other well is at $|110\rangle$
(site 7)  and $|111\rangle$ (site 8). (a) The theoretical
predictions; (b) The liquid NMR experimental results. The eight bars
at each step indicate the particle's probabilities on the states
$|000\rangle$, $|001\rangle$, $|010\rangle$,
$|011\rangle$, $|100\rangle$, $|101\rangle$, $|110\rangle$, and
$|111\rangle$, from left to right. The initial state is
prepared to be $|110\rangle$. The bars with slashes (red and brown) stand for the probabilities in the well of $|110\rangle$ and $|111\rangle$. The bars with dots (blue and green) stand for the probabilities in the well of $|010\rangle$ and $|011\rangle$. As is illustrated in (a) and (b), the
particle tunnels from the well in sites 3 and 4 to the other well. The oscillations within each well are also observed.}
\label{three-qubitresult}
\end{figure}
%

%
\begin{figure}[tbp]
\begin{center}
\includegraphics[width=180mm,height=88mm]{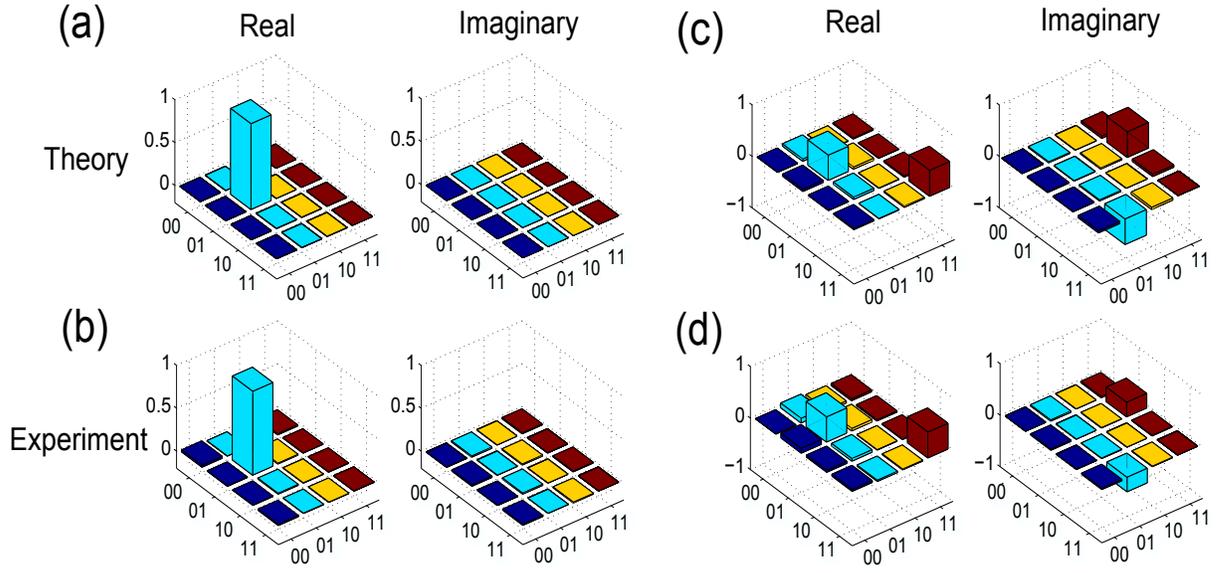}
\end{center}
\caption{
The real and imaginary parts of (a) the density matrix
elements of the $|01\rangle$ state; (b) the experimentally reconstructed density matrix elements for the initial $|01\rangle$ state in our two-qubit
experiments; (c) the theoretically
calculated density matrix elements after nine steps of evolution
with a double-well potential; (d) the experimentally reconstructed density matrix elements after the nine steps of evolution with a double-well potential in our two-qubit
experiments.}
\label{aresult}
\end{figure}

%
\begin{figure}[tbp]
\begin{center}
\includegraphics[width=180mm]{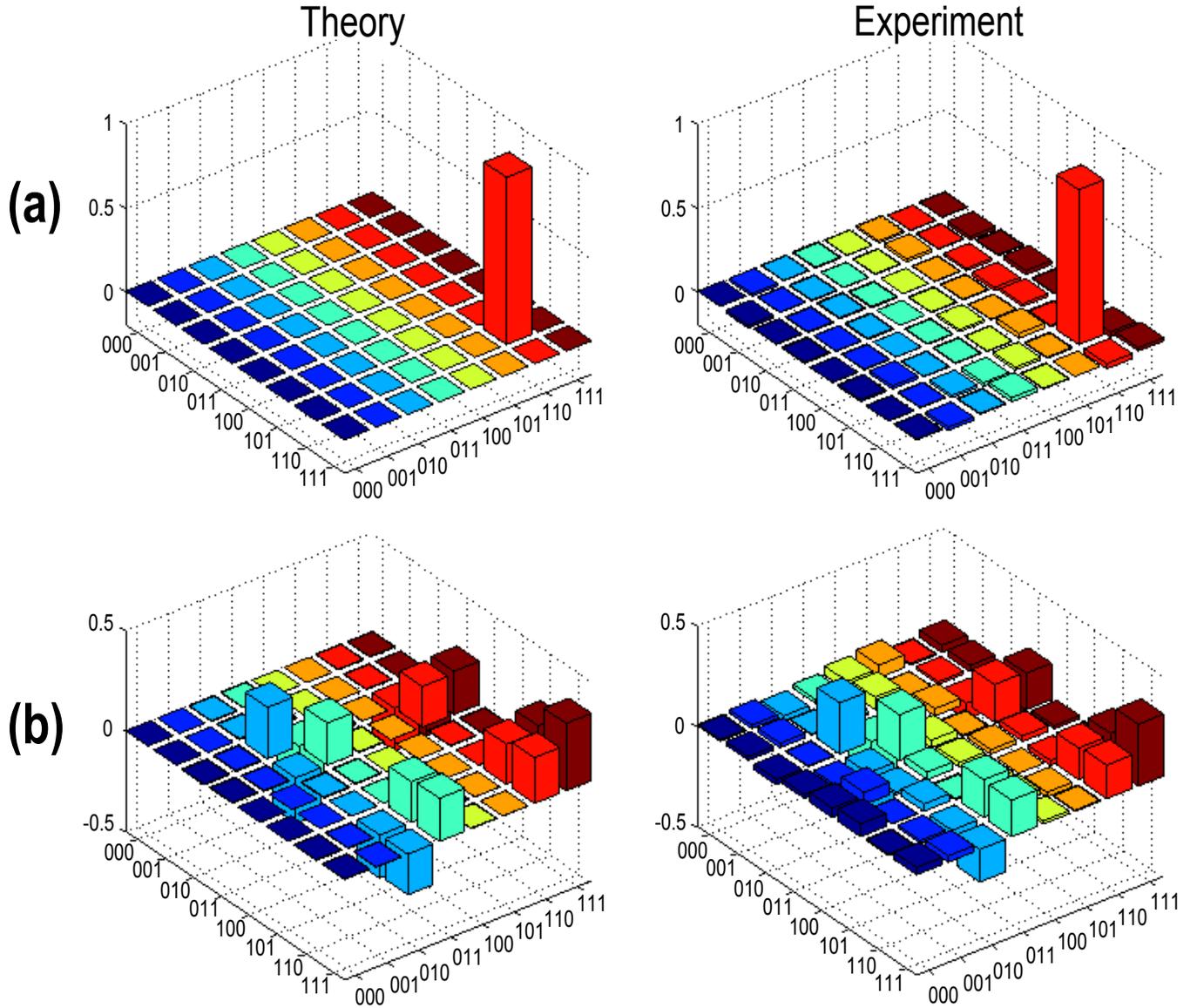}
\end{center}
\caption{
The real parts of (a) the theoretically
calculated and experimentally reconstructed density matrix elements for the initial $|110\rangle$ state in our three-qubit
experiments; (b) the theoretically
calculated and experimentally reconstructed density matrix
elements after the five time steps of evolution with a double-well potential in our three-qubit
experiments.}
\label{finalstate}
\end{figure}

\begin{figure}[tbp]
\begin{center}
\includegraphics[width=88mm]{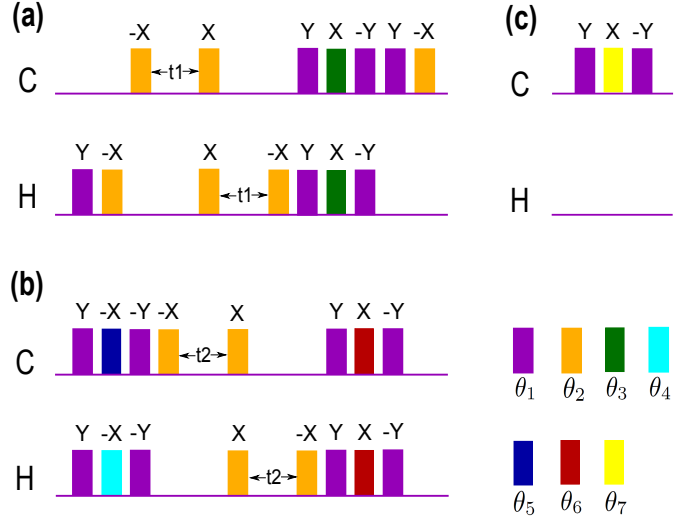}
\end{center}
\caption{
Pulse sequences to implement the circuit in Fig. 1,
where (a), (b), (c) realize the operations $F$, $D$,
and $Q$ respectively. The time periods $t1=\frac{1}{8J}=580.9\mu s$
and $t2=\frac{\pi}{40J}=365.0\mu s$ represent the free evolution
durations under $J$-coupling . The
bars represent single-qubit rotation pulses of different angles, and the corresponding rotation axes are labeled above them. The seven different colors indicate different
rotation angles as $\theta_{1}=\frac{\pi}{2}$, $\theta_{2}=\pi$,
$\theta_{3}=\frac{\pi}{4}$, $\theta_{4}=\frac{\pi^{2}}{40}$, $\theta_{5}=\frac{\pi^{2}}{10}$,
$\theta_{6}=\frac{\pi^{2}}{20}$, $\theta_{7}=2$, respectively.}
\label{sequence}
\end{figure}

\end{document}